\documentclass[aip,reprint]{revtex4-2}
\usepackage{amsfonts,amsmath,bm,amssymb,color}
\usepackage{graphicx}
\usepackage[makeroom]{cancel}

\usepackage{rotating}
\usepackage{tikz,tkz-euclide}
\usetikzlibrary{decorations.pathmorphing}
\usetikzlibrary{arrows}
\usetikzlibrary{arrows.meta}
\usetikzlibrary{decorations.markings}
\usepackage{pgfplots}
\usepackage{tkz-fct}
\usetikzlibrary{calc,patterns}

\usepackage[unicode=true]{hyperref}

\definecolor{bostonuniversityred}{rgb}{0.8, 0.0, 0.0}
\definecolor{dukeblue}{rgb}{0.0, 0.0, 0.61}
\definecolor{ao(english)}{rgb}{0.0, 0.5, 0.0}
\definecolor{darkmagenta}{rgb}{0.55, 0.0, 0.55}
\definecolor{armygreen}{rgb}{0.29, 0.33, 0.13}
\definecolor{coquelicot}{rgb}{1.0, 0.22, 0.0}
\definecolor{fucsiak}{rgb}{0.4, 0.08, 0.4}
\definecolor{airforceblue}{rgb}{0.36, 0.54, 0.66}

\begin{document}

\title{Velocity structure function in a geostrophic coherent vortex under strong rotation}
\author{Leon L. Ogorodnikov}
\email{leon.ogorodnikov@hse.ru}
\author{Sergey S. Vergeles}
\email{ssver@itp.ac.ru}

\affiliation{Landau Institute for Theoretical Physics, Russian Academy of Sciences,\\1-A Akademika Semenova av., 142432 Chernogolovka, Russia}
\affiliation{National Research University Higher School of Economics, Faculty of Physics,\\Myasnitskaya 20, 101000 Moscow, Russia}

\begin{abstract}
     We consider analytically pair structure function of turbulent pulsations on the background of a coherent geostrophic vortex in a fast rotating fluid. The statistics of the turbulent pulsation is determined by their dynamics which is the dynamics of inertia waves affected by the differential rotation in the vortex and weak viscous damping. Our consideration is restricted by the smallest scales, where the velocity field remains smooth. We establish the anisotropy of the structure function. The velocity gradient of the turbulent pulsations achieves its largest value for the radial direction and its smallest value in the streamwise direction, resembling its behaviour in turbulent flow with mean shear component without imposed rotation.
\end{abstract}

\maketitle

\section{Introduction}

Correlation functions of the velocity pulsations are the basic tool for the characterization of the spatial structure of the flows which have turbulent or chaotic component. The most known limiting cases of the turbulent flow is statistically isotropic turbulence of incompressible fluid in three \cite{frisch1995turbulence} and two \cite{tabeling2002two} dimensions. It is usually that the turbulence is superimposed on a mean steady flow. A particular but important case of the mean flow is a shear flow. It is realized near walls, especially in pipe flow\cite{mullin2011experimental}. In a shear flow, the amplitude of the velocity in turbulent pulsations is of the order of the mean flow. Geometrically, the turbulent pulsations are highly elongated in streamwise direction \cite{biferale2005anisotropy}, and the velocity is preferably directed along the elongation direction.  The turbulent pulsations of the largest sizes take kinetic energy from the mean flow and then give is to smaller scales, that is the onset for the direct energy cascade. Despite the assumption that the turbulent pulsations should become statistically isotropic at small scales, only second and third order correlation functions correspond to isotropic case whereas high-order structure functions at the small scales inherit the anisotropy of the large-scale shear flow.

In two-dimensional turbulent flows\cite{laurie2014universal} and in three-dimensional rotating turbulent flows large-scale mean flow can become coherent. This means in particular, that its influence on small-scale turbulent pulsations become strongly dominative, so the statistics of the small-scale flow is dramatically different from the isotropic case. The main case of the coherent flow is an axially symmetric vortex. In the vortex, the mean flow is a differential rotation which locally is a shear flow. Structure function in two-dimensional coherent vortex was analytically calculated in Ref.~\onlinecite{kolokolov2016velocity}

In three-dimensions, the fast global rotation splits the full flow onto slow two-dimensional geostrophic flow and fast three-dimensional inertia waves. Under definite conditions, the geostrophic flow can be developed into a coherent vortex\cite{kolokolov2020structure}. The differential rotation inside the vortex is locally three-dimensional shear flow which rotates around spanwise direction. This treatment brings together the consideration of the statistics of the small-scale turbulent pulsations inside the vortex with the same problem for flow in a pipe which rotates fast around spanwise direction\cite{yang2020mean}.

In this paper, we analytically find the structure function of small-scale turbulent pulsations inside a coherent geostrophic vortex. We restrict ourselves by the consideration of small distances, where the velocity field is smooth. We assume that the Rossby number for the large-scale flow is small, and the shear rate dominates both the nonlinear self-interaction and the viscous dissipation rate for the inertia waves. Besides the statistics of the velocity pulsations is interesting itself, it determines passive scalar advection and mixing at smallest scales, if the Schmidt (or Prandtl) number, which is the ratio of the kinematic viscosity of the fluid and the diffusion coefficient of the scalar is large\cite{antonia2003effect}.

%The mixing of passive scalar in a turbulent flow with strong shear component is a long-standing problem. For example,  Passive scalar transport in rotating channel flow\cite{brethouwer2018passive}. Passive scalar diffusion in two-dimension in non-smooth velocity field Kraichnan model \cite{celani2005shear}

\section{General properties of pair correlation function}

We consider a flow ${\bm v}(t,{\bm r})$ which contains a mean component ${\bm U}$ and a turbulent part ${\bm u}$, ${\bm v}={\bm U}+{\bm u}$. It is assumed that the mean component ${\bm U}$ changes at times and distances much larger than the characteristic time and space scales of ${\bm u}$. The temporal mean of the turbulent part of the flow is zero, $\langle {\bm u}\rangle=0$. The object of our investigation is the simultaneous structure function $S^{ij}({\bm r})$ of the velocity turbulent pulsations. In the limit of strong scale separation, the statistics of the turbulent pulsations can be adopted to be locally homogeneous in space. Then the structure function depends on only the separation distance ${\bm r}$,
\begin{equation}\label{structure-function-definition}
    S^{ij}({\bm r})
    =
    \langle \big(u^j({\bm r}^\prime+{\bm r})
        -
        u^j({\bm r}^\prime)\big)
        \big(u^i({\bm r}^\prime+{\bm r})
        -
        u^i({\bm r}^\prime)\big)
        \rangle,
\end{equation}
where the averaging is assumed to be either over time or the position ${\bm r}^\prime$. The region of the space averaging should be less than the characteristic scale of the mean flow ${\bm U}$. The structure function is symmetric by definition, $S^{ij}({\bm r}) = S^{ji}({\bm r}) = S^{ij}(-{\bm r})$.

We are considering turbulence in a rotation fluid with angular velocity ${\boldsymbol \Omega}$ and denote the direction of the rotation ${\bf e}_z = {\boldsymbol \Omega}/|{\boldsymbol \Omega}|$. Here a cartesian reference system $Oxyz$ is assumed to be determined. In what follows we assume that the pulsating part ${\bm u}$ is relatively weak, so its self-nonlinear influence can be neglected as compared with the its nonlinear interaction with the mean velocity ${\bm U}$. Then the dynamics of the turbulent part ${\bm u}$ is described by the Navier-Stokes equation linearized with respect to ${\bm u}$. The global rotation is strong, so large-scale Rossby number is small, ${\mathrm Ro}_{\scriptscriptstyle R}\sim \nabla {\bm U}/2\Omega\ll1$. Then it is natural to expand the velocity field over plane waves with circular polarizations,
\begin{equation}\label{appendix:02}
    {\boldsymbol v}(t,{\bm r})
    \ = \
    \sum_{s=\pm1}\int(\mathrm{d}^3k)
    \,a_{{\bf k}s}(t)\,{\bf h}_{\bf k}^s\,
    \exp\big(i({\bf k}\!\cdot\!{\bm r})\big),
\end{equation}
where the basis vectors ${\bf h}_{\bf k}^s =\big([{\bf k}\times[{\bf k}\times {\bf e}_z]]+ isk[{\bf k}\times {\bf e}_z]\big) /{\sqrt{2}k^2\sin\theta_{\bf k}}$ are normalized to unity. The dynamics of expansion coefficients $a_{{\bf k}s}$ is determined by the dispersion law of inertia waves\cite{greenspan1968theory}. As Rossby number for the large-scale flow is assumed to be small, the inertial waves with the opposite circular polarizations are uncorrelated \cite{kolokolov2020structure}. The homogeneity of the statistics in space means that the square mean of the expansion coefficients is
\begin{equation}
    \langle a_{{\bf q}\sigma}\, a_{{\bf k}s}\rangle
    =
    (2\pi)^3\,\delta({\bf q}+{\bf k}) K_{{\bf k}}\delta_{\sigma s},
\end{equation}
where the mean kinetic energy  stored in the turbulent pulsations with wavevector ${\bf k}$ satisfies the symmetry $K_{{\bf k}} = K_{-{\bf k}}$. Here we consider the case when the helicity is absent, that is the wave statistics is symmetric under transformation $s\to-s$. As a result, the velocity structure function (\ref{structure-function-definition}) is parametrized by one scalar function:
\begin{equation}\label{Fij-simple}
    S^{ij}({\bm r})
    =
    2\hat \delta_{\scriptscriptstyle \perp}^{ij}\left(  K({\bm 0})-K({\bm r}) \right) ,
    \quad
    K({\bm r}) = \int \frac{\mathrm{d}^3k}{(2\pi)^3}
    K_{\bf k}e^{i{\bf k}{\bm r}},
\end{equation}
where $\hat \delta_{\scriptscriptstyle \perp}^{ij}$ is a linear integral operator in ${\bm r}$-space which Fourier transform is the transverse projector $\delta_{\scriptscriptstyle \perp}^{ij} = \delta^{ij} - k^ik^j/{\bf k}^2$.

\section{Rapid distortion theory in columnar vortex}

Now consider a coherent geostrophic vortex. We choose a cylindrical coordinates $\{\rho,\varphi,z\}$, which axis coincides with the axis if the vortex. The large-scale velocity ${\bm U}$ has the only nonzero azimuth component $U(\rho)$ in the case. To develop rapid distortion theory\cite{batchelor1954effect,kolokolov2020structure,parfenyev2021velocity} for the description of the turbulent pulsation dynamics, we choose a local Cartesian reference system $\{\xi,\eta,z\}$, which origin moves with azimuth velocity $U(\rho_0)$ at distance $\rho=\rho_0$ from the vortex axis. In addition, the reference system rotates with the angular velocity $U(\rho_0)/\rho_0$. In the reference system, the large-scale velocity field is constant in time and locally is a shear flow with rate $\Sigma(\rho_0)$, where $\Sigma = \rho\partial_\rho(U/\rho)$. Axis $\xi$ is directed streamwise. Axis $\eta$ is directed toward the vortex axis, so it is the shear velocity gradient-direction axis. We call the direction \textquoteleft radial\textquoteright\ instead the common \textquoteleft normal\textquoteright\ when considering a shear flow near a wall. Direction of $z$-axis can be called \textquoteleft spanwise\textquoteright. The equation describing the linearized dynamics of small-scale turbulent pulsations can be derived\cite{kolokolov2020structure} in wavevector space. According to the equation, the pulsation $a_{{\bf k}s}$ having wavevector ${\bf k}$ is an inertia wave which oscillates with frequency $\omega_{\bf k} = 2\Omega (k^z/k)$. The inertia wave is influenced by the large-scale flow due to its weak inhomogeneity in space. Small large-scale Rossby number ${\mathop{\mathrm{Ro}}}_{\scriptscriptstyle R}=\Sigma/2\Omega\ll1$ leads to the dynamics of inertia waves with the same wavevector ${\bf k}$ and opposite polarizations $s$ are uncoupled. Advection by the inhomogeneous large-scale flow leads to the wave evolution occurs along characteristics
\begin{equation}\label{ROT:03}
    {\bf k}^\prime(t)
    =
    \{k^\xi,k^\eta+\mathop{\mathrm{sign}}\Sigma\cdot t k^\xi,k^z\},
\end{equation}
where $t$ is dimensionless time that is time measured in $1/|\Sigma|$.

Further, we assume that there is some external random source which excites the turbulent part of the flow. Physically, the origin of the force can be buoyancy force which arises due to advection of heat. The assumed absence of mean spirality of the flow means, in particular, that the force produces zero spirality. The statistics of the force is assumed to be homogeneous in space, its mean power per unit mass is $\epsilon$. Its correlation function in the Fourier space is characterized by the function $\chi({\bf k})$, which is normalized to unity, $\int \chi({\bf k})\,\mathrm{d}^3k/(2\pi)^3 = 1$. We will assume that it is isotropic and decays at $k\sim k_f$ and trends to zero at $k\to 0$. In the case, the source of the anisotropy of the velocity correlation function is only the anisotropy of the large scale velocity field. The viscosity influence for the turbulent pulsations is assumed to be weak, so dimensionless parameter $\gamma = \nu k_f^2/\Sigma \ll 1$.

\section{Structure function}

Let the wavevector of a turbulent pulsation is equal to ${\bf k}$ in the moment of excitation $-\tau$ by the external force \cite{kolokolov2016structure,kolokolov2020structure}. At the time of measurement, the wavevector of the wave is ${\bf k}^\prime(\tau)$ (\ref{ROT:03}). Then the aggregated contribution from all the preceding times and all the wavevectors into the structure function is
\begin{eqnarray}\nonumber
    S^{ij}({\bm r})
    &=&
    \frac{\epsilon}{|\Sigma|}
    \int\limits_0^{+\infty} {\mathrm d}\tau
    \int\limits
    \frac{{\mathrm d}^3k}{(2\pi)^3}
    \frac{k\chi({\bf k})}{k^\prime}
    \,\delta_{{\scriptscriptstyle \perp}}^{\prime ij}
    \left(1-e^{i{\bf k}^\prime {\bm r}}\right)
    e^{-\Gamma},
    \\[5pt] &&\label{structfunc00}
    \Gamma
        =
        \frac{2\gamma}{|\Sigma|}\int\limits_0^{\tau}
        {\mathrm d}\varsigma \,
        k^{\prime 2}(\varsigma),
\end{eqnarray}
where $\delta_{\scriptscriptstyle \perp}^{\prime ij}=\delta_{\scriptscriptstyle \perp}^{ij}\big({\bf k}^\prime\big)$ and ${\bf k}^\prime$ without argument is ${\bf k}^\prime(\tau)$. In the paper, we consider small distances, for which the velocity field can be approximated by linear function in space. Then it is sufficient to expand the structure function up to the second order in ${\bm r}$:
\begin{equation}\label{structfunc}
    S^{ij}({\bm r})
    =
    \frac{\epsilon r^mr^n}{|\Sigma|}
    \int\limits_0^{+\infty} {\mathrm d}\tau
    \int\limits
    \frac{{\mathrm d}^3k}{(2\pi)^3}
    k\chi({\bf k})
    \,
    \frac{\delta_{{\scriptscriptstyle \perp}}^{\prime ij}
        k^{\prime m} k^{\prime n}}
        {k^\prime}
    e^{-\Gamma}.
\end{equation}
In (\ref{structfunc}), one can  first perform integration over wavevectors and after that over inverse time $\tau$. At large times, $\tau \gg 1$, and for typical values $k^\xi\lesssim1$, the movement along characteristics sweeps wavevectors out of region $k\lesssim 1$ as $k^\prime(\tau)\sim \tau\gg 1$. Law (\ref{ROT:03}) means, that the corresponding contribution into (\ref{structfunc}) is suppressed by factor $1/\tau$. Additionally, the viscosity leads to the exponential suppression, that becomes relevant if $\tau>\tau_{\ast} = \gamma^{-1/3}$. The viscous exponent in (\ref{structfunc}) can be evaluated as $\Gamma \sim (\tau/\tau_{\ast})^3/3$ when $k^\xi\sim1$, but it is reduced to less than unity if
\begin{equation}\label{kxi-confinement}
    |k^{\xi}|\lesssim (\tau_\ast/\tau)^{3/2}
\end{equation}
and while $\tau < 1/\gamma$. Below we call times $\tau<\tau_{\ast}$ \textquoteleft ballistic times\textquoteright, and times $\tau_{\ast} <\tau<1/\gamma$ \textquoteleft viscous times\textquoteright.

Consider first the trace of the structure function $S = S^{ii}({\bm r})$. As $\delta_{{\scriptscriptstyle \perp}}^{\prime ii}=2$, the convergence of integral (\ref{structfunc}) is determined by the ratio $k^{\prime m}k^{\prime n}/k^\prime$ and the viscous exponent $\Gamma$ (\ref{structfunc00}). We start from $\eta^2$-contribution into (\ref{structfunc}), when $n=m=\eta$. At large ballistic times, $|k^\xi|\sim 1$ so the ratio $(k^{\prime \eta})^2/k^\prime\sim \tau |k^\xi|$ and the time integrand in (\ref{structfunc}) is proportional to $\tau$. At viscous times, the viscous exponent imposes the confinement (\ref{kxi-confinement}), so $(k^{\prime \eta})^2/k^\prime \sim \sqrt{\tau_{\ast}^3/\tau}$. As the result, the time integrand is proportional to $\tau_{\ast}^3/\tau^2$. Thus, the time integral (\ref{structfunc}) is determined by times $\tau\sim\tau_{\ast}$ and the coefficient before $\eta^2$ is $\sim \tau_{\ast}^2$.

Next, consider $\xi^2$-contribution. At large times, the ratio $(k^\xi)^2/k^\prime\sim k^\xi/\tau$. Thus, the integrand in (\ref{structfunc}) is proportional to $1/\tau$ that leads to logarithmic divergency at times $\tau \sim\tau^*$. At viscous times, the integrand decreases faster due to confinement (\ref{kxi-confinement}). Thus, the coefficient before $\xi^2$ is $\sim \ln \tau_{\ast}$.

For $z^2$-contribution, the ratio $(k^z)^2/k^\prime\sim (k^z)^2/\sqrt{\tau^2(k^\xi)^2+(k^z)^2}$. There is a logarithmic divergency at $|k^\xi|>1/\tau$ in integral (\ref{structfunc}), which has a cutoffs $k^\xi\sim 1$ at ballistic times and (\ref{kxi-confinement}) at viscous times. Thus, the time integrand in (\ref{structfunc}) is $\sim\tau^{-1}\ln\tau$ at ballistic times and \hbox{$\sim \tau^{-1}\ln\sqrt{\tau_{\ast}^{3}/\tau}$} at viscous times. As a result, the coefficient before $z^2$ is \hbox{$\sim \ln^2\tau_{\ast}$}. The last nonzero contribution is for \hbox{$m=\xi$}, \hbox{$n=\eta$}. The corresponding ratio $k^\xi k^{\prime \eta}/k^\prime\sim |k^\xi|$. Accounting the same cutoff for $|k^\xi|$, one finds the coefficient \hbox{$\sim \tau_{\ast}/\Sigma$} including sign in $\xi\eta$-term.

The overall result is
\begin{equation}\label{01}
    S^{ii}({\bm r})
    \sim
    \frac{\epsilon k_f^2}{|\Sigma|}
    \left(\ln\tau_{\ast} \cdot\xi^2 +
        (\tau_{\ast} \eta +\mathop{\mathrm{sign}}\Sigma\cdot\xi)^2 +
        \ln^2\tau_{\ast} \cdot z^2\right).
\end{equation}
Here numerical coefficients are not identified, as they depend also on the specific parameters of the external force correlation function $\chi({\bf k})$. The result should be compared with one-point mean $\langle {\bm u}^2\rangle\sim (\epsilon/|\Sigma|)\ln^2\gamma$ \cite{kolokolov2020structure}. The direction of the fastest increase of the structure function is the radial direction. More precisely, the direction deviates from the radial direction at angle \hbox{$\sim1/\tau_{\ast}$} toward the streamwise direction. Accordingly, the direction of the slowest increase of the structure function deviates from the streamwise direction at the same angle. The angle scales with the shear rate and viscosity as $(\nu k_f^2/\Sigma)^{1/3}$, that is characteristic scaling law for the advection-diffusion process in a stationary shear flow \cite{ranz1979applications}. Note that since $\delta_{{\scriptscriptstyle \perp}}^{\prime \xi\xi}\approx \delta_{{\scriptscriptstyle \perp}}^{\prime zz}$ at times $\tau\gg1$,
\begin{equation}\label{02}
    S^{\xi\xi} = S^{zz} = S^{ii}/2.
\end{equation}

Now consider $S^{\eta\eta}$, which turns out to be smaller than $S^{\xi\xi}$. For $\eta^2$-contribution, $(k^{\prime \eta})^2$ multiplier cancels the denominator in $\delta^{\prime\eta\eta}_{\scriptscriptstyle \perp} = (k^{\xi2}+k^{z2})/k^{\prime 2}$ and further calculations repeat those for $z^2$-contribution into $S^{ii}$. For $z^2$-contribution, integration of $\delta^{\prime\eta\eta}_{\scriptscriptstyle \perp}$ with multiplier $1/k^\prime$ over $k^\xi$ gives $1/\tau$ at ballistic times $\tau\gg1$. The time integrand decreases faster at viscous times.
Thus the coefficient before $z^2$ is $\sim \ln\tau_{\ast}$. For $\xi^2$-contribution, $\delta^{\prime\eta\eta}_{\scriptscriptstyle \perp}$ being integrated over $k^\xi$ with multiplier $(k^\xi)^2/k^\prime$ leads to integrand $\tau^{-3}\ln\tau$ at large times $\tau\gg1$. Thus, the coefficient \hbox{$\sim1$} before $\xi^2$ is determined by times $\tau\sim1$. Finally, the time-dependent part of the integration in (\ref{structfunc}) for $\xi\eta$-contribution is determined by $\tau\sim1$. Finally, the time-dependent part of the integration in (\ref{structfunc}) for $\xi\eta$-contribution is determined by $\tau\sim1$. It can be calculated with integration by parts,
\begin{eqnarray}\label{by-parts}
    \mathop{\mathrm{sign}}\Sigma\cdot
    \int\limits_0^\infty \mathrm{d}\tau
    \frac{k^{\prime \eta}k^\xi}{k^{\prime 3}} e^{-\Gamma}
    =
    \frac{1}{k}
    -
    \gamma\int\limits_0^\infty \mathrm{d}\tau k^\prime e^{-\Gamma}.
\end{eqnarray}
The overall answer is
\begin{eqnarray}
    S^{\eta\eta}
    \sim
    \frac{\epsilon k_f^2}{|\Sigma|}
    \left(
    \xi^2 + \eta\xi + \ln^2\tau_{\ast}\cdot \eta^2 + \ln\tau_{\ast} \cdot z^2
    \right),
\end{eqnarray}
where again the numerical coefficients are not identified.

The calculations of the coefficients before $\xi^2$ and $z^2$ in $S^{\xi\eta}$ involves the same way (\ref{by-parts}). Calculation of the coefficients before $\eta^2$ and $\xi\eta$ repeats those for the $\xi\eta$- and $\xi^2$-term in $S^{ii}$. Collecting all the terms, we arrive to
\begin{equation}
    S^{\xi\eta}
    \sim
    -\frac{\epsilon k_f^2}{|\Sigma|}
    \left(\xi^2 +
        \ln\tau_{\ast}\cdot\xi\eta
        +
        \tau_{\ast} \eta^2
        +
        z^2
        \right)
\end{equation}
In the similar manner, one can find the remaining nonzero matrix elements of the structure function:
\begin{equation}
    S^{\xi z}
    \sim
    -\frac{\epsilon k_f^2}{|\Sigma|}
    \eta z,
    \qquad
    S^{\eta z}
    \sim
    -\frac{\epsilon k_f^2}{|\Sigma|}
    \left(
        \xi z
        +
        \ln^2\tau_{\ast}\cdot\eta z
        \right).
\end{equation}

\section{Conclusion}

To summarize, we have established the properties of the structure function of the turbulent pulsations inside a geostrophic coherent vortex at the smallest scales, where the velocity field of the pulsations can be approximated by the linear profile in space. The main property of the velocity statistics at the scales is that the streamwise and the spanwise $\xi,z$-components of the velocity changes in radial (shear-increment) $\eta$-direction faster in $\tau_\ast = (\Sigma/\nu k_f^2)^{1/3}$ times than in streamwise and spanwise directions, see (\ref{01},\ref{02}). The structure function of $\eta$-component is logarithmically smaller and does not contain the spatial anisotropy.

Large dimensionless parameter $\tau_\ast$ corresponds to passive scalar mixing in a stationary shear flow \cite{ranz1979applications}. Although the dynamics of the turbulent pulsations in a rotating fluid is primarily fast oscillations of inertia waves, the velocity structure function does not depend on the angular velocity $\Omega$, if the coherent vortex is large enough, so its radius $R\gg \tau_\ast/(\Omega k_f)$. Otherwise, one should reduce $\tau_\ast$ by the redefinition $\tau_\ast \sim 1/(\Omega k_f R)$.

\section{Acknowledgments}

This work was supported by the Russian Science Foundation, Grant No. 20-12-00383. S.S.V. acknowledges the support from Grant No. 19-1-2-46-1 of the Foundation for the Advancement of Theoretical Physics and Mathematics “BASIS.”

%aipnum4-2.bst 2019-01-14 (MD) hand-edited version of apsrev4-1.bst
%Control: key (0)
%Control: author (8) initials jnrlst
%Control: editor formatted (1) identically to author
%Control: production of article title (0) allowed
%Control: page (1) range
%Control: year (1) truncated
%Control: production of eprint (0) enabled
%

\end{document}